\documentclass[journal]{IEEEtran}
\ifCLASSINFOpdf
\else
   \usepackage[dvips]{graphicx}
\fi
\usepackage{url}
\usepackage{graphicx}

\usepackage{graphicx}
\usepackage{amssymb}
\usepackage{amsmath}
\usepackage{algorithm}
\usepackage{algpseudocode}
\usepackage{xcolor}
\usepackage{enumitem}
\usepackage[labelformat=simple]{subcaption}

\usepackage{scalerel}

\usepackage[style=numeric,firstinits,url=false,sorting=none,maxbibnames=9]{biblatex}

\usepackage[font=small]{caption}

\renewbibmacro{in:}{
  \ifentrytype{article}
    {}
    {\bibstring{in}
     \printunit{\intitlepunct}}}

\begin{document}

\title{Conservative Quantum Offline Model-Based Optimization}
\author{Kristian Sotirov, 
Annie E. Paine,
Savvas Varsamopoulos,
Antonio A. Gentile,
Osvaldo Simeone, \IEEEmembership{Fellow, IEEE}

\thanks{Kristian Sotirov is with the Department of Engineering, King's College London, WC2R2LS, London, U.K. (email: kristian.sotirov@kcl.ac.uk).  Annie E. Paine, Savvas Versamopoulos, and Antonio A. Gentile are with Pasqal SAS, 24 Av. Emile Baudot, 91120 Palaiseau, France (email: annie.paine@pasqal.com; savvas.varsamopoulos@pasqal.com; andrea.gentile@pasqal.com). Osvaldo Simeone is with the Institute for Intelligent Networked Systems  (INSI),  Northeastern University London, One Portsoken Street, London E1 8PH, United Kingdom   (email: o.simeone@northeastern.edu).}

\thanks{The work of K. Sotirov was supported by King's College London and Pasqal through the King’s Quantum Centre for Doctoral Training. The work of O. Simeone was supported by the European Research Council (ERC) under the European Union’s Horizon Europe Programme (grant agreement No. 101198347), by an Open Fellowship of the EPSRC (EP/W024101/1), and by the EPSRC project (EP/X011852/1).}
}

\maketitle

\begin{abstract}
    Offline model-based optimization (MBO) refers to the task of optimizing a black-box objective function using only a fixed set of prior input-output data, without any active experimentation. Recent work has introduced quantum extremal learning (QEL), which leverages the expressive power of variational quantum circuits to learn accurate surrogate functions by training on a few data points. However, predictive models may incorrectly extrapolate objective values in unexplored regions, leading to the selection of overly optimistic solutions. In this paper, we propose integrating QEL with conservative objective models (COM), a regularization technique aimed at ensuring cautious predictions on out-of-distribution inputs.  
    {\color{black} Empirical results on synthetic benchmark optimization tasks demonstrate that the resulting hybrid algorithm, COM-QEL, reliably finds solutions offering a better trade-off between novelty and usefulness.}
\end{abstract}

{\color{black}
\begin{IEEEkeywords}
    Model-based optimization, quantum neural networks, quantum optimization, quantum extremal learning
\end{IEEEkeywords}
}

\section{Introduction} \label{introduction}
    
\subsubsection{Context and Motivation}
\emph{Offline model-based optimization} (MBO) is an important primitive in science and engineering. The goal of MBO is to identify configurations that maximize a \emph{black-box} objective function using only a static dataset of prior evaluations \cite{kim2025offline}. Unlike online optimization, which permits interactive queries or experiments, the offline setting does not allow for the collection of any new data due to cost or risk constraints. This scenario arises in many high-stakes applications. For example, one may design a molecule with desired properties using only existing experimental data \cite{pmlr-v97-brookes19a}, or design a novel aircraft with optimal characteristics based on existing prototypes \cite{hoburg2014geometric}. In all these cases, online querying the real objective function is prohibitively expensive or even infeasible, so the optimization must rely solely on the precomputed dataset.

\begin{figure}
\begin{center}

    \begin{subfigure}[t]{0.5\textwidth}
        \includegraphics[scale=0.64]{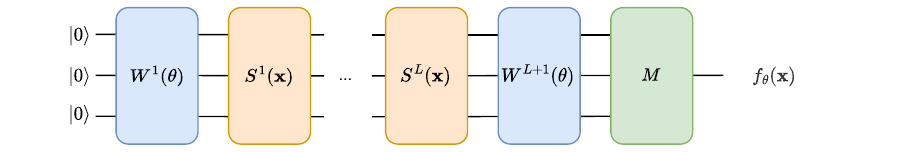}
        
        \captionsetup{justification=centering}
        \hspace{2.5cm}
        \vspace{-0.5cm}
        \caption{}
        \label{fig:pqc}
    \end{subfigure}
    \hspace{-0.5cm}
    \begin{subfigure}[t]{0.5\textwidth}
        \hspace{-0.5cm}
        \includegraphics[scale=0.45]{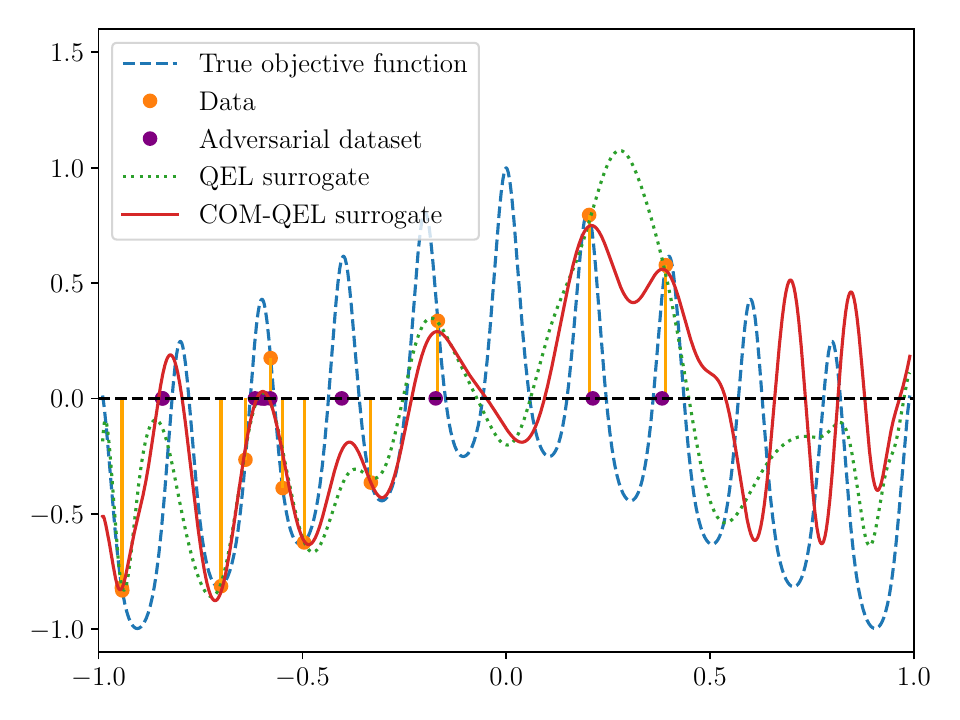}
        \centering
        \vspace{-0.3cm}
        \caption{}
        \label{fig:example}
    \end{subfigure}
    \caption{(a) QEL \cite{QEL} uses a parameterized quantum circuit (PQC) as a surrogate $f_\theta(x)$ for an unknown black-box function $f(x)$ using real data $\mathcal{D}$. (b) COM-QEL extends QEL by augmenting the surrogate model design objective with a regularizer that penalizes high values on adversarial, out-of-sample inputs.}
    \vspace{-0.7cm}
\end{center}
\end{figure}

However, offline MBO poses significant challenges compared to its online counterpart. The primary difficulty is \emph{extrapolation uncertainty}: the true objective values for inputs not present in the dataset are unknown and can deviate substantially from model predictions. A learned model might erroneously predict very high objective values in under-sampled regions, enticing the optimizer toward those regions. This phenomenon, often referred to as \emph{model exploitation} or \emph{objective value hacking}, can lead to selecting designs that appear optimal under the model but perform poorly in reality \cite{kim2025offline}.

\emph{Quantum extremal learning} (QEL), proposed in \cite{QEL}, is a quantum algorithm that can be used effectively for offline MBO. QEL employs a parameterized quantum circuit (PQC)  \cite{Schuld:2021mml,simeone2022introduction}  as a surrogate model that is variationally trained on the available data (see Fig. \ref{fig:pqc}). 
Differentiation of the quantum circuit against the encoded input-dependence is then leveraged to estimate the value of the input variable that extremizes the learned model. 
QEL illustrates the potential of quantum models to perform optimization tasks, possibly offering advantages in expressiveness or in navigating complex objective landscapes. 

However, the original QEL method did not specifically incorporate mechanisms to protect against model overestimation on unseen inputs (see Fig. \ref{fig:example} for an example). This form of regularization towards more conservative surrogate models has proven beneficial in classical offline MBO \cite{pmlr-v139-trabucco21a,yu2021roma,qi2022data,dao2024incorporating}. Thus, there remains a gap in adapting quantum optimization algorithms to the stringent demands of the offline setting, motivating this work.

\subsubsection{Related Work}

Classical approaches to offline MBO have been extensively studied in recent years \cite{kim2025offline}. The most common strategy is to train a \emph{surrogate model} on the dataset that approximates the objective function, and then optimize the surrogate while accounting for the model’s uncertainty on unseen inputs \cite{NIPS2012_05311655,pmlr-v80-garnelo18a}. An alternative family of classical methods pursues generative modeling to directly propose candidate solutions \cite{kumar2020model}.

A fundamental challenge for both surrogate and generative strategies in offline MBO is ensuring conservative predictions outside the support of the data. To address this,  \cite{pmlr-v139-trabucco21a} introduced \emph{conservative objective models} (COM), which impose a penalty on the surrogate model’s predicted objective value for out-of-sample inputs. COM adds regularization terms during model training that lower the predictions on unseen or uncertain regions, effectively encouraging the model to be pessimistic about areas not backed by data.

Another recent work \cite{grudzien2024functional} leverages known structural properties of the objective function via \emph{functional graphical models} (FGM) to constrain the optimization, demonstrating that incorporating problem-specific knowledge can further enhance offline data-driven optimization.

\subsubsection{Main Contributions}

In this work, we develop \emph{quantum extremal learning with conservative objective models} (COM-QEL), a novel algorithm that synergizes the strengths of QEL and COM for improved offline optimization. The main contributions of this paper are summarized as follows:
\begin{itemize}[leftmargin=*]

\item	\emph{Integration of conservative modeling into quantum surrogates}: We generalize the QEL algorithm \cite{QEL} by including a penalty that drive down predictions on inputs outside the support of the training data, thereby aligning the quantum model with the principles of offline conservatism \cite{pmlr-v139-trabucco21a}.

\item	\emph{Structured quantum surrogates via functional graphical models}: We integrate the FGM structure \cite{grudzien2024functional} of the underlying objective function, into QEL by leveraging a quantum graph neural network (QGNN) ansatz \cite{verdon2019quantum}.

 \item	\emph{Empirical performance gains}: Through experimental evaluations of benchmark offline optimization tasks (see Fig. \ref{fig:example}), we demonstrate that COM-QEL consistently outperforms the original QEL algorithm in terms of the trade-off between usefulness and novelty \cite{kim2025offline}. Furthermore, we show that structure-aware QGNN ansatz can improve the usefulness of the solution obtained with both QEL and COM-QEL. 
\end{itemize}

\section{Background} \label{background}

As illustrated in Fig. \ref{fig:example}, in offline MBO, we are given a dataset $\mathcal{D}$ containing $N$ input-output pairs $\mathcal{D}= \{\mathbf{x}_i, y_i\}_{i = 1}^N$, where the output $y_i = f(\mathbf{x}_i)$ for an unknown real-valued objective function $f(\mathbf{x})$ with continuous convex domain $\mathcal{X} \in \mathbb{R}^d$. Our goal is to find an input value $\mathbf{x}^* \in \mathcal{X}$, that approximates an optimal solution for the objective function $f(\mathbf{x})$, i.e.,
\begin{equation}\label{eq:optim}
    \mathbf{x}^* \in \underset{\mathbf{x} \in \mathcal{X}}{\arg\max} \ f(\mathbf{x}).
\end{equation} 

As discussed in Sec. \ref{introduction}, a common class of MBO methods trains a surrogate function $f_\mathbf{\theta}(\mathbf{x})$ to serve as a model for the underlying function $f(\mathbf{x})$. This is done by optimizing the parameters to fit the dataset $\mathcal{D}$. In particular, QEL \cite{QEL} uses a PQC as a surrogate model. Accordingly, as shown in Fig. \ref{fig:pqc}, the surrogate function is given by 
\begin{equation}\label{eq:surrogate}
f_\theta(\mathbf{x}) = \left  \langle 0 \right | U_\theta^{\dagger}(\mathbf{x})M U_\theta(\mathbf{x})\left  | 0 \right \rangle,
\end{equation}

\noindent where $|0\rangle$ is the initial state of the system of $n$ qubits, $U_\theta(\mathbf{x})$ represents the $2^n \times 2^n$ unitary matrix implemented by the PQC and $M$ is a $2^n \times 2^n$ observable matrix. 

A common layout for the trainable section of a PQC -- i.e. the \emph{ansatz} -- applies a sequence of parameterized unitary matrices $W^{l}(\theta)$ and data encoding unitary matrices $S^l(\mathbf{x})$ across $L$ layers indexed as $l = 1, \dots L$, with a final parameterized unitary matrix $W^{L+1}(\theta)$ \cite{PhysRevA.103.032430}, yielding the unitary matrix 
\begin{equation}\label{eq:data-reuploading}
U_\theta(\mathbf{x}) = W^{L+1}(\theta)S^L(\mathbf{x})W^{L}(\theta)\dots S^1(\mathbf{x})W^{1}(\theta).
\end{equation}

\noindent The unitary matrices $\{W^{l}(\theta)\}_{l = 1}^{L+1}$ and $\{S^{l}(\mathbf{x})\}_{l = 1}^{L}$ typically consist of sequence of parameterized single-input gates and fixed two-qubit gates \cite{Schuld:2021mml,simeone2022introduction}.

The PQC parameters $\theta$ are optimized by minimizing the mean squared loss between
 the true output values $y_i = f(\mathbf{x}_i)$ and the predictions $f_\mathbf{\theta}(\mathbf{x}_i)$ made by the surrogate model over the training examples $(\mathbf{x}_i, y_i)$ in the dataset $\mathcal{D}$:  
\begin{equation}\label{eq:qelmin}
    \mathbf{\theta}^\textrm{QEL} = \underset{\theta}{\arg \min} \frac{1}{N}\sum_{i = 1}^N(f_\theta(\mathbf{x}_i) - y_i)^2.
\end{equation}

\noindent The minimization in (\ref{eq:qelmin}) is typically carried out using perturbation-based gradient estimates via parameter-shift rules \cite{PhysRevA.98.032309, PhysRevA.99.032331}.

Once the surrogate model $f_{\theta^\textrm{QEL}}(\mathbf{x})$ is obtained, a solution $\textbf{x}^{\textrm{QEL}} \in \mathcal{X}$ approximating the maximum in (\ref{eq:optim}) is evaluated via gradient ascent starting from the best solution in the dataset $\mathcal{D}$. Define as $\mathbf{x}_{\textrm{max}}$, with $(\mathbf{x}_{\textrm{max}}, y_\textrm{max}) \in \mathcal{D}$ and where $y_\textrm{max}  = 
\max \{y_i: (\mathbf{x}_i, y_i) \in \mathcal{D}\}$ the best solution in the dataset. Applying gradient ascent on the surrogate model, QEL yields the sequence of iterates 
\begin{equation}\label{eq:optim-input}
    \mathbf{x}^{t} = \mathbf{x}^{t-1} + \mu^t\nabla_\mathbf{x}f_{\theta^{\textrm{QEL}}}(\mathbf{x}^{t-1}),  
\end{equation}
 \noindent for $t = 1,\dots,T$, where $\mathbf{x}^0 = \mathbf{x}_{\textrm{max}}$, the number of iterations is denoted as $T$, and $\mu^t > 0$ is a sequence of learning rates. If the domain $\mathcal{X}$ is restricted, i.e., if $\mathcal{X} \subset \mathbb{R}^d$, a projection step is applied after each iteration \eqref{eq:optim-input}. Alternatively, one can also use \emph{reflective}  methods that return iterates within the interior of the domain $\mathcal{X}$ (see next section) \cite{sato2025convergence}.  Finally, QEL returns the solution $\mathbf{x}^\textrm{QEL} = \mathbf{x}^{T}$.

\section{Quantum Extremal Learning with Conservative Objective Models} \label{method}

In this section, we introduce COM-QEL, a novel approach that extends QEL \cite{QEL} by leveraging the regularization technique introduced for classical MBO in  \cite{pmlr-v139-trabucco21a}. The goal is to create a more conservative surrogate model, reducing the risk of 
erroneously overestimating values of the underlying function $f(x)$ away from the sampled inputs (see Fig. \ref{fig:example}).

\subsubsection{Surrogate Training via Adversarial Regularization}

Conservative regularization aims at minimizing the values of the surrogate function $f_\theta(\mathbf{x})$ at adversarial inputs chosen from the set encountered during the optimization of the surrogate function. 
{\color{black}
To elaborate, let $\mathbf{x}_{\theta, T_p}(\mathbf{x}_0)$ denote an adversarial input obtained from a true data point $\mathbf{x}_0 \in \mathcal{D}$ by applying $T_p$ gradient ascent steps \eqref{eq:optim-input} using the current surrogate function $f_\theta(\mathbf{x})$. The set of adversarial inputs is then given as $\mathcal{D}_{\theta,T_p} = \{\mathbf{x}_{\theta, T_p}(\mathbf{x}) \}_{\mathbf{x} \in \mathcal{D}}$,  including all inputs obtained from the training set $\mathcal{D}$.
}

Formally, COM-QEL addresses the constrained problem
\begin{equation}\label{eq:tau}
\begin{split}
    \theta^{\textrm{COM-QEL}} = &\underset{\theta}{\arg \min} \frac{1}{N}\sum_{i = 1}^N(f_\mathbf{\theta}(\mathbf{x}_i) - y_i)^2\\
    &\mathrm{s.t.} \ \frac{1}{N}\sum_{i = 1}^N f_\theta(\mathbf{x}_{\theta, T_p}(\mathbf{x}_i)) - \frac{1}{N}\sum_{i = 1}^N f_\theta(\mathbf{x}_i)\leq \tau,
\end{split}
\end{equation}
 where $\tau > 0$ is a hyperparameter. The constraint in (\ref{eq:tau}) requires that the values of the surrogate function on the adversarial inputs do not exceed, on average, the values of the surrogate function on the inputs $\{\mathbf{x}_i\}_{i =1}^N$ in the dataset $\mathcal{D}$ by more than a threshold $\tau$.

The constrained problem \eqref{eq:tau} is transformed into the unconstrained problem 
\begin{equation}\label{eq:unconstrained}
    \min_\theta \max_{\alpha \geq 0} \left \{\mathcal{L}(\theta, \alpha) = \frac{1}{N}\sum_{i = 1}^N(f_\mathbf{\theta}(\mathbf{x}_i) - y_i)^2 + \alpha C(\mathbf{x}, \theta ) \right \},
\end{equation}

\noindent where $C(\mathbf{x}, \theta ) = \frac{1}{N}\sum_{i = 1}^N f_\theta(\mathbf{x}_{\theta, T_p}(\mathbf{x}_i)) - \frac{1}{N}\sum_{i = 1}^N f_\theta(\mathbf{x}_i) - \tau$ and $\alpha \geq 0$ is a hyperparameter that controls the conservatism of the model. The problem \eqref{eq:unconstrained} is then addressed via dual gradient descent-ascent, with a primal descent step with respect to parameters $\theta$ and a dual ascent step with respect to $\alpha$ \cite{nesterov2009primal}.

Having obtained the PQC parameters $\theta^{\textrm{COM-QEL}}$ through the solution of the problem  (\ref{eq:unconstrained}), COM-QEL obtains the solution $\mathbf{x}_{\theta^{\textrm{COM-QEL}}}(\mathbf{x}_{\textrm{max}})$ using gradient ascent as in (\ref{eq:optim-input}) with function $f_{\theta^{\text{COM-QEL}}}(\mathbf{x})$.

\subsubsection{Implementation Details}

{\color{black}

Assuming a closed and bounded domain $\mathcal{X}$, each entry of the input vector $\mathbf{x}$ is normalized to take values in the interval $[-1, 1]$.  With this choice, a reflective gradient ascent step \eqref{eq:optim-input} is implemented entry-wise: after computing the tentative gradient step \eqref{eq:optim-input},  we check whether the new point $x^t$ is within the range $[-1, 1]$, and if not, the gradient is modified via the reflective step $x^t=2c-(x^{t-1}+\mu^t \partial f_{\theta}(\mathbf{x}^{t-1})/\partial x)$, with $c = 1$ if the gradient \eqref{eq:optim-input} exceeds the upper bound $1$ and $c = -1$ otherwise.

This approach is applied for both the main optimization procedure and the adversarial dataset generation.  
}

To generate adversarial samples $\mathbf{x}_{\theta, T_p}(\mathbf{x}_0)$, we use a single gradient ascent step, i.e., $T_p = 1$, with learning rate $\mu^t = 0.05\sqrt{d}$, where $d$ is the dimension of the input sample $\mathbf{x}$ \cite{pmlr-v139-trabucco21a}. We run the dual gradient ascent optimization for problem (\ref{eq:tau}) for 100 epochs using the Adam optimizer with learning rate 0.05 \cite{kingma2017adammethodstochasticoptimization}. Gradients are evaluated using the parameter shift rule \cite{PhysRevA.98.032309, }.

\section{Numerical Results and Conclusions}\label{results}

In this section, we compare the performance of COM-QEL against  QEL \cite{QEL} and COM 
\cite{pmlr-v139-trabucco21a} in synthetic benchmarks involving continuous optimization domains \cite{kim2025offline}, considering first unstructured functions and then a structured graph-based function. 

\subsubsection{Performance Measures}

As evaluation metrics, we use the novelty and the usefulness score functions \cite{kim2025offline}.
The usefulness score measures the improvement of a solution $\hat{\mathbf{x}}$ as compared to the best solution in the dataset $\mathcal{D}$, i.e.,

\begin{equation}
U(\hat{\mathbf{x}})= \frac{f({\hat{\mathbf{x}}})-y_{\min}}{y_{\max}-y_{\min}},
\end{equation}

\noindent where $y_\textrm{min}  = 
\min \{y_i: (\mathbf{x}_i, y_i) \in \mathcal{D}\}$ and $y_\textrm{max}  = 
\max \{y_i: (\mathbf{x}_i, y_i) \in \mathcal{D}\}$ are the worst and best values of the objective function $f(\mathbf{x})$ for solutions within the dataset $\mathcal{D}$, respectively.
The novelty score of an outcome $\hat{\mathbf{x}}$ measures how different the output $\hat{\mathbf{x}}$ is in relation to the existing dataset $\mathcal{D}$. i.e.,\begin{equation}
N(\hat{\mathbf{x}})= \min_{\mathbf{x}\in \mathcal{D}} ||\hat{\mathbf{x}}-\mathbf{x}||^2.
\end{equation}

\subsubsection{Ansatz}

For tasks involving unstructured functions, we adopt the ansatz (\ref{eq:data-reuploading}) in Fig. \ref{fig:pqc} \cite{PhysRevA.103.032430}, where each layer consists of a hardware-efficient ansatz (HEA) $W^{l}(\theta)$  and of an encoding circuit $S^{l}(\mathbf{x})$, applied in this order \cite{Kandala_2017}. The HEA encompasses generic single-qubit gates implemented as the cascade of $X$, $Z$, and $X$ Pauli rotations \cite{nielsen2010quantum}, which are applied in parallel to all $n$ qubits, and of two-qubit CNOT gates, where qubit $i$ is the control qubit and qubit $i + 1$ is the target qubit for all $i = 1, \dots, n-1 $. For the encoding block, we use Chebyshev tower encoding \cite{kyriienko2021solving,PhysRevA.103.032430}: given input $x$, the Pauli $Y$ rotation $\hat{R}_y(2j\arccos(x))$ is applied to neuron $i$ in layer $l$ with
$j = (l-1)n + i$. 
 The surrogate $f_\theta(\mathbf{x})$ in (\ref{eq:surrogate}) is finally obtained as the expectation value of the observable $ M = \sum_{i = 1}^nZ_i$, where $Z_i$ is the Pauli $Z$ matrix operating on the $i$-th qubit.

For the classical neural network used to implement the original COM algorithm \cite{pmlr-v139-trabucco21a}, we employ a feedforward neural network with a single hidden layer. The size of the hidden layer is chosen so that the number of weights in the classical neural network matches the number of parameters in the quantum neural network. This approach enables a direct comparison of classical and quantum models under a common constraint (see, e.g., \cite{alcazar2020classical}). 

\subsubsection{Low-Bandwidth Objective Function}

{ \color{black} As a first example, we consider the two-dimensional function  $f(\mathbf{x}) = \sum_{i = 1}^2\cos(2\pi x_i) (1 - 0.1|x_i|)$,  with domain $x_1, x_2\in [-1, 1]$ \cite{kim2020benchmark}. For this low-bandwidth function,  we consider a PQC with $n = 4$ qubits, where the first two qubits encode variable $x_1$ and the last two qubits encode $x_2$, and $L = 3$ layers. The results for the usefulness and novelty are presented in Fig. \ref{fig:cosine_usefulness-novelty} via violin plots obtained by randomly and uniformly sampling $N = 20$ data points for the dataset $\mathcal{D}$.  For COM-QEL, we provide results with different values of hyperparameter $\tau$.

\begin{figure}
\begin{center}
    \includegraphics[width=0.9\columnwidth]{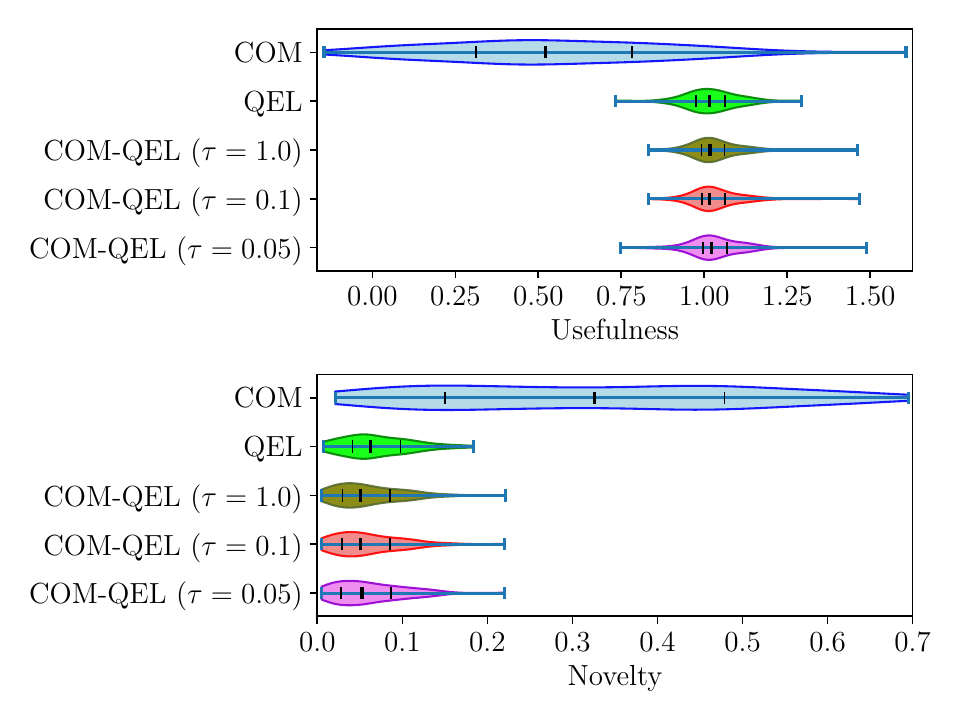}
    \vspace{-0.3cm}
    \caption{{\color{black} Usefulness (top) and novelty (bottom) for classical COM \cite{pmlr-v139-trabucco21a}, QEL \cite{QEL} and COM-QEL for  $\tau = 0.05, \tau = 0.1$ and $\tau = 1$ over 100 different sets of $N=20$ randomly generated points in the domain $[-1, 1]^2$, for a two-dimensional function.}}
    \label{fig:cosine_usefulness-novelty}
    \vspace{-0.5cm}
\end{center}
\end{figure}

The figure illustrates how the conservative objective modeling adopted by COM-QEL allows the derivation of better solutions compared to QEL, while avoiding solutions with excessively low usefulness, as long as hyperparameter $\tau$ is well chosen. In this regard, it is observed that COM-QEL is quite robust to the choice of hyperparameter $\tau$, with minor variations in usefulness distribution for the range of values between $\tau=0.1$ and $\tau=1$. The improved utility of COM-QEL is achieved by selecting more conservative solutions on average, compared to QEL.  In addition, although classical COM can potentially find more novel solutions, most of the solutions have lower usefulness compared to the worst possible outcomes for both QEL and COM-QEL.} 

\subsubsection{High-Bandwidth Objective Function}

\begin{figure}
\begin{center}
    \includegraphics[width=0.9\columnwidth]{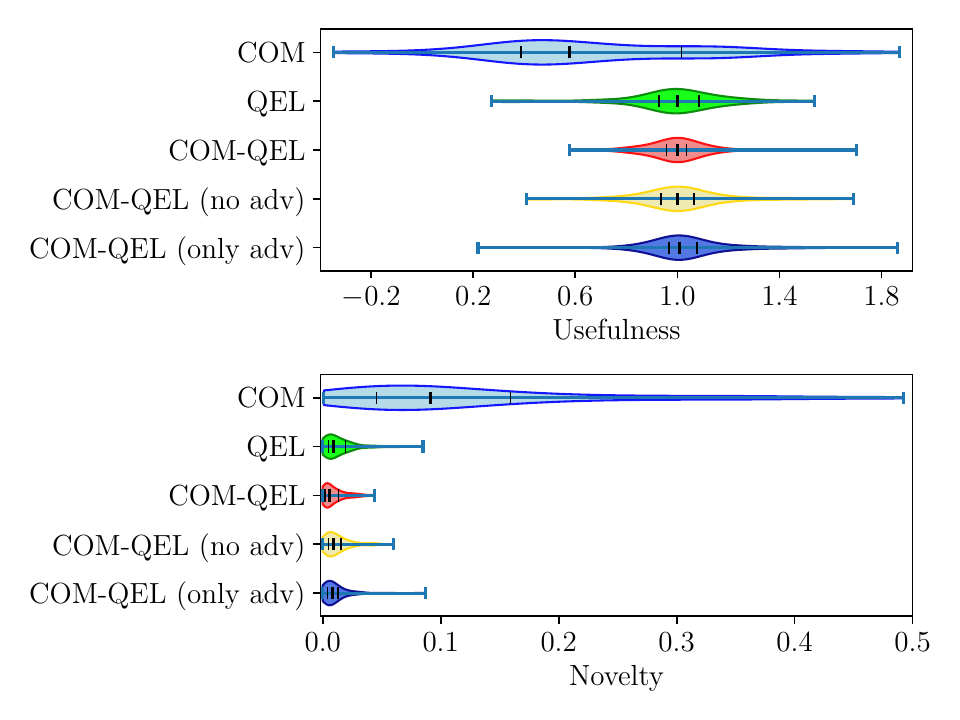}
    \vspace{-0.3cm}
        \caption{ \color{black} {Usefulness (top) and novelty (bottom) for classical COM \cite{pmlr-v139-trabucco21a}, QEL \cite{QEL},   COM-QEL, and two COM-QEL variants applying only one of the two penalty terms in (\ref{eq:unconstrained})  over 100 different sets of $N = 10$ randomly generated points in the domain $[-1, 1]$, for the 1D Ackley function.}}
    \label{fig:ackley1_usefulness-novelty}
    \vspace{-0.5cm}
\end{center}
\end{figure}

{\color{black} Consider now a more challenging synthetic function with large fluctuations, namely the Ackley function \cite{kim2020benchmark}. For this scalar function, we chose $n = 3$ qubits, all encoding the single input $x \in [-1, 1]$.  As shown in Fig. \ref{fig:ackley1_usefulness-novelty}, COM-QEL can avoid solutions with low usefulness levels. In this challenging example, this goal is accomplished by choosing solutions with a lower novelty as compared to the other models. To demonstrate the role of the full penalty term in (\ref{eq:unconstrained}), we also show the performance of two variants of COM-QEL: the first  incorporates only the first penalty term in $C(\mathbf{x}, \theta)$, which accounts only  for the adversarial dataset $\mathcal{D}_{\theta, T_p}$ (``only adv.''), while the second includes only the other term in $C(\mathbf{x}, \theta)$, depending on the initial set $\mathcal{D}$ (``no adv.''). The results confirm the importance of retaining both penalty terms in order to enhance usefulness, while ensuring sufficient novelty.
}

\subsubsection{Structured Functions}

Consider now a structured function $f(x_1,x_2,x_3)=f_a(x_1,x_2)+f_b(x_3)$, where $f_a(x_1,x_2)=100(x_2 - x_1^2) + (x_1 - 1)^2 $ is the two-dimensional Rosenbrock function \cite{kim2025offline} and $f_b(x_3)$ is the Ackley function considered in Fig. \ref{fig:ackley1_usefulness-novelty}. Using the terminology in \cite{grudzien2024functional}, the two sets of variables  $\{x_1, x_2\}$ and $\{x_3\}$ are functionally independent, as they appear in different terms of the objective function. Graphically, the two sets form two separate cliques of the underlying  FGM (see Sec. I).

\begin{figure}
\begin{center}
    \includegraphics[width=0.9\columnwidth]{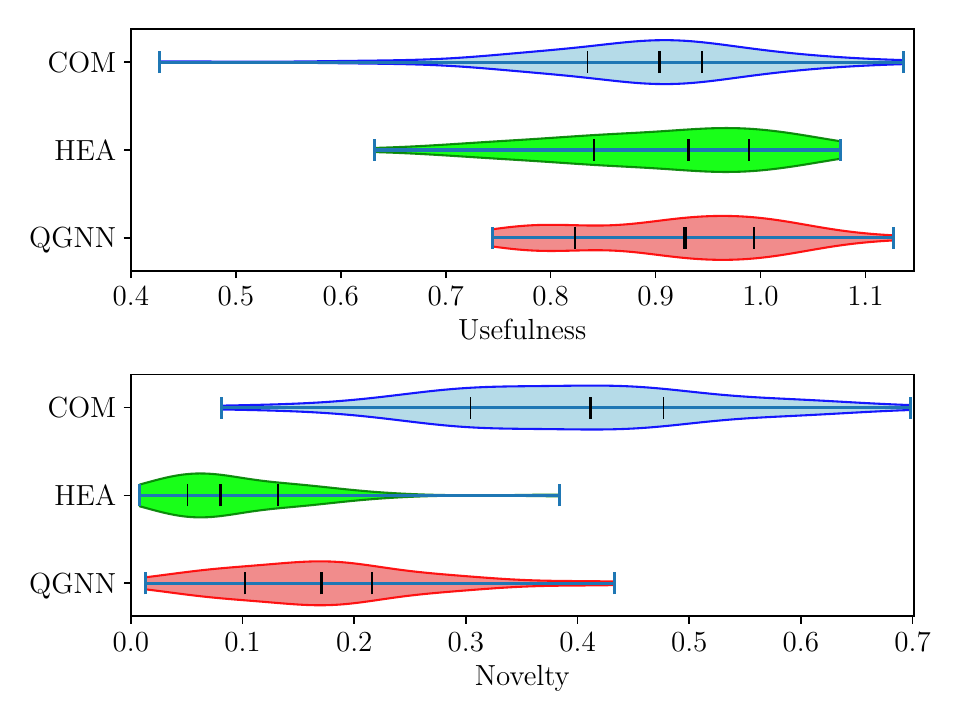}
        \vspace{-0.3cm}
        \caption{Usefulness (top) and novelty (bottom) for classical COM \cite{pmlr-v139-trabucco21a}, COM-QEL (HEA) and COM-QEL (QGNN)  over 50 different sets of $N = 30$ randomly generated points in the domain $[-1, 1]^3$ for a structured function.} 
    \label{fig:ackley1-rosenbrock_usefulness-novelty}
    \vspace{-0.5cm}
\end{center}
\end{figure}

The PQC used in this example contains $L = 6$ layers and $n = 6$ qubits, where the first two qubits encode variable $x_1$, the next two qubits encode $x_2$, and the last two qubits encode $x_3$. Apart from the HEA adopted earlier, to capture the special structure of this objective function, we also consider a QGNN-like ansatz \cite{verdon2019quantum}, in which two-qubit CNOT gates are added only between qubits that are in the same clique of the FGM.

The violin plots in Fig. \ref{fig:ackley1-rosenbrock_usefulness-novelty} compare COM with COM-QEL implemented using either the HEA or the QGNN, after $50$ runs with $N = 30$ randomly and uniformly sampled data points. The figure shows that COM-QEL with the QGNN ansatz outperforms COM-QEL with the HEA in terms of both usefulness and novelty. Furthermore, although the QGNN ansatz obtains less novel predictions compared to classical COM, it yields solutions with a similar maximum usefulness, while safeguarding against solutions with extremely low usefulness.

{ \color{black} \subsubsection{Conclusions} Overall, our results  show that COM-QEL has the ability to effectively explore out-of-sample input regions for better-behaved functions,  while generally refraining from moving too far away from the dataset for more challenging, highly varying functions with many local optima. 
Moreover, the performance of COM-QEL can be improved by encoding the underlying problem structure into the quantum circuit. Future work could explore experimental validations on real quantum devices, as well as the extension to  offline reinforcement learning.
}

\printbibliography

\end{document}